\documentclass[acmtog,screen,nonacm]{acmart}

\AtBeginDocument{
  }

\setcopyright{none}
\usepackage{booktabs}
\usepackage{amsmath}
\usepackage{amsfonts}
\usepackage{nicefrac}
\usepackage{xcolor}
\usepackage{epigraph}
\usepackage{algorithm}
\usepackage{algpseudocode}

\title[Before the Shutter: Aesthetic and Actionable Portrait Photography Planning in 3D Scenes]{Before the Shutter: Aesthetic and Actionable Portrait Photography Planning in 3D Scenes}

\author{\href{https://j-rx.com}{Ruixiang Jiang}}
\email{rui-x.jiang@connect.polyu.hk}
\affiliation{
  \institution{The Hong Kong Polytechnic University}
  \city{Hong Kong SAR}
  \country{China}
}

\author{\href{https://chenlab.comp.polyu.edu.hk/}{Chang Wen Chen}}
\email{chen.changwen@polyu.edu.hk}
\affiliation{
  \institution{The Hong Kong Polytechnic University}
  \city{Hong Kong SAR}
  \country{China}
}

\renewcommand{\shortauthors}{Jiang and Chen}
\authorsaddresses{}

\begin{document}

\begin{abstract}
Portrait photography is largely decided before the shutter opens: the subject's pose, the camera configuration, and the lighting devices must be coordinated within the surrounding 3D scene. In contrast, most existing computational methods focus on post-production in 2D image space, such as retouching, relighting, or editing images that already exist; pre-capture photographic planning remains largely unexplored. We introduce 3D aesthetic portrait planning, the task of generating human pose, camera, lighting, and exposure plans that produce visually compelling portraits while satisfying geometric and photometric feasibility in a 3D scene. Our approach builds a Photographic Scene Graph that represents scene affordances, subject-scene relations, and portrait-relevant lighting structure. Built on this representation, we perform aesthetic-guided comparative planning over previous attempts and current viewfinder observations. Experiments across diverse indoor and outdoor scenes show that our method produces portraits preferred by human raters and MLLM evaluators over competitive baselines, while maintaining high physical plausibility. Together, our results suggest a path from post-capture correction toward pre-capture computational portrait planning. Project repository: \url{https://github.com/songrise/Before-the-Shutter}.
\end{abstract} 

\begin{teaserfigure}
    \includegraphics[width=\textwidth]{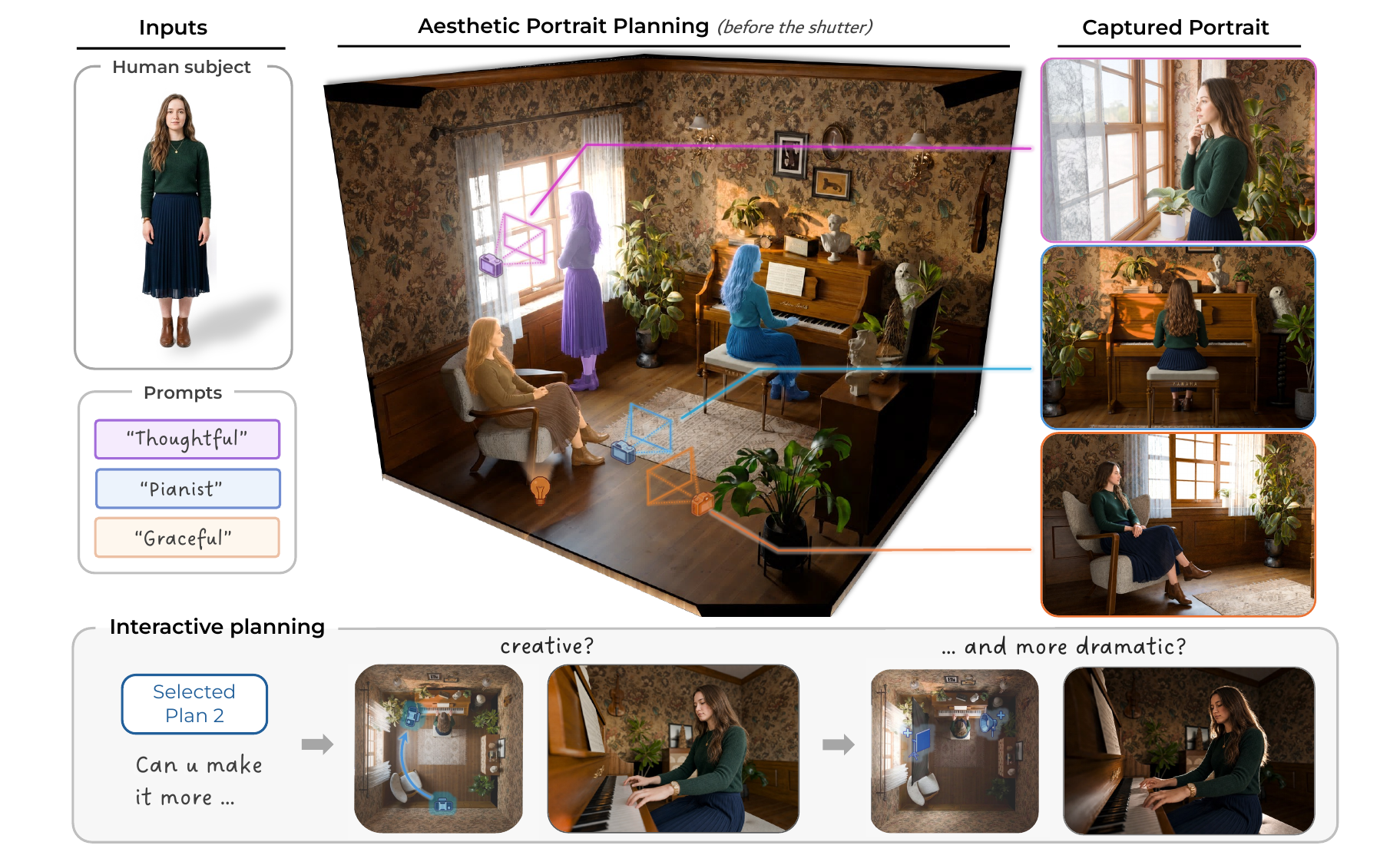}
    \caption{Given a 3D scene, a human subject, and user prompts, our system generates candidate portrait plans before capture by jointly exploring subject pose, camera placement, lighting, and exposure in 3D space. Each plan is visualized in the scene and linked to its rendered viewfinder image. Iterative prompting and comparative feedback refine the plan frontier toward the desired photographic intent. }
    \label{fig:teaser}
\vspace{0.5cm}
\end{teaserfigure}
\maketitle

\section{Introduction}
\epigraph{``You don't take a photograph, you make it.''}{Ansel Adams}

Portrait photography is one of the most universal and enduring forms of visual expression. From professional studio sessions to casual smartphone snapshots, capturing a person in a way that is readable and appealing remains a widely shared aspiration. Advances in camera hardware and computational photography have made cameras more capable than ever: a sharp, well-exposed image is now trivial to achieve with a single click. Yet the problem of making an aesthetically compelling portrait, where the pose communicates, the light sculpts, and the composition feels considered, remains as challenging as ever.

To understand what \emph{makes} a portrait, we can separate the photographic
process into pre-shutter planning and post-shutter refinement~\cite{adams1981negative,hunter2015light} stages. Given a scene and a human actor, pre-shutter planning organizes the relationship among the subject,
the camera, and the lighting to aesthetically communicate the desired effect. The difficulty lies in the fact
that these choices are both coupled and scene-dependent: pose, camera, lens,
exposure, and lighting each change how the others read, and the same setup can
succeed in one environment but fail in another. This abundance of interacting
choices is the expressive power of pre-shutter planning: for a trained
photographer, pose, viewpoint, lens, and light become instruments for aesthetic
intent. At the same time, knowing how to balance these choices is precisely what
makes portrait photography difficult.

Despite the importance of this pre-shutter planning, most computational approaches to portraiture begin only after an image has been captured. While paradigms such as retouching~\cite{su2025styleretoucher}, relighting~\cite{rao2024lite2relight}, re-composition through cropping~\cite{zhong2021aestheticcrop}, and general image editing~\cite{wanimage2026} demonstrate their power in 2D refinement, they are structurally confined to 2D image space: they cannot access the 3D degrees of freedom that govern pre-shutter decisions, and therefore cannot produce physically actionable plans. Such reactive approaches may suggest \emph{what} a more aesthetic portrait could look like, but they do not answer \emph{how} to make it. For example, a retouching method may brighten a shadowed face, but it does not tell us where to place the key and fill lights, or how to set their relative power, to avoid flat or uncontrolled lighting. A cropping method may improve 2D composition, but it cannot choose the camera position and focal length needed to
preserve background context while keeping the subject visually balanced. All of these important pre-shutter planning decisions remain outside the scope of current systems. This leaves a gap between image-space enhancement and the scene-level decisions that photographers actually control before capture.

In this paper, we ask whether these pre-shutter decisions can be modeled computationally. Given a static 3D scene, a human subject, and a user prompt specifying the desired portrait style, our goal is to generate a set of candidate portrait plans -- each specifying a subject pose and placement, camera configuration, controllable lighting, and exposure -- that are both visually compelling and physically actionable in the scene. We call this task 3D aesthetic portrait planning. It is an inverse problem by nature: aesthetic quality is observed only in the resulting 2D viewfinder image, yet the available controls are 3D and tightly coupled. This asymmetry makes feed-forward prompt-to-plan prediction brittle: the aesthetic effect of pose, camera, and light often becomes clear only through scene-grounded interaction and observation. We therefore propose a planning algorithm that simultaneously understands the spatial and photometric structure of the scene, and iteratively refines a frontier of candidate plans through comparative aesthetic judgement.

Our contributions are fourfold. First, we formulate \textbf{aesthetic portrait planning} as prompt-conditioned generation of subject pose, camera, lighting, and exposure plans, and introduce a benchmark of 50 tasks spanning 14 diverse indoor and outdoor scenes. Second, we propose a \textbf{Photographic 
Scene Graph} that jointly encodes spatial affordances, subject-scene relations, and portrait-relevant lighting structure, enabled grounded portrait planning. Third, we introduce \textbf{Aesthetic-Guided Comparative Planning}, a training-free algorithm that maintains an aesthetic 
frontier of candidate plans and refines it through comparative aesthetic signal, supporting both MLLM-as-Judge for fully automatic planning and 
human judgement for interactive refinement. Fourth, we instantiate these 
components with coordinated \textsc{Photographer}, \textsc{Actor}, and \textsc{Judge} roles and show that it improves human and MLLM aesthetic preference over image-only, one-pass, greedy, and spatial-graph baselines while maintaining high actionability.

\section{Related Work}

\subsection{Computational portrait aesthetics.}
Computational portrait methods, including retouching~\cite{chang2025pertouch,elezabi2024inretouch}, relighting~\cite{sun2019single,pandey2021total}, and composition~\cite{yuan2024clipcropping}, operate primarily in 2D image space on captured images, which makes them post-shutter by design. To improve attractiveness, these methods use Image Aesthetic Assessment (IAA) models~\cite{murray2012ava,schuhmann2022laionaesthetics,kirstain2023pickapic} as aesthetic proxies. However, canonical IAA score is not interpretable and usually provide limited guidance for converting an image-level aesthetic preference into concrete 3D changes in pose, camera, or lighting. Recent work suggests that MLLMs encode human-aligned and interpretable aesthetic priors than IAA models~\cite{jiang2025mllmaesthetics}, motivating pairwise comparison as a more flexible planning signal than absolute scoring. We instead propose a training-free planning approach supporting both MLLM-as-Judge and human judge, where an aesthetic frontier is refined through pairwise comparative judgement. This facilitates translation of image aesthetic signal into actionable plans in 3D scene.

\subsection{Scene-aware human placement and interaction.}

Human-scene interaction (HSI) methods recover or synthesize plausible human bodies in 3D environments by reasoning over geometry, semantic affordances, contact relations, and physical constraints~\cite{savva2016pigraphs,hassan2019resolving,zhang2020place,zhao2022compositional,li2024genzi}. These methods ground the body in the surrounding scene -- chairs afford sitting, walls afford leaning, contacts constrain stable pose -- but are evaluated purely on physical plausibility. This is insufficient for portrait photography in three respects: a physically stable pose may be visually illegible from the camera, poorly lit by available illumination, or unexpressive under the intended prompt. Our work inherits physical grounding from this literature as an abstract \textsc{Actor} role, which is coordinated with the \textsc{Photographer} and \textsc{Judge} to jointly satisfy camera legibility, illumination, and aesthetic preference.

\subsection{Camera control and portrait lighting design.}
Camera placement and lighting are fundamental creative controls in computer
graphics, but they have largely been studied in isolation from each other and from subject pose. Intelligent camera control and virtual cinematography methods plan
viewpoints in 3D scenes under constraints such as visibility, framing, shot size, composition, continuity, and aesthetics~\cite{halper2000camplan,
christie2008camera,lino2011computational,galvane2015continuity, xie2023gait, zhang2025gendop, liu2024chatcam}. These systems
show that camera placement is an expressive scene-level decision, but they
usually assume that actor pose and illumination are fixed. Portrait relighting and reflectance-capture methods approach the problem from
the opposite direction, demonstrating the importance of light direction,
softness, and subject-background separation~\cite{
nestmeyer2020learning, pandey2021total, mei2024holorelight}, yet most of them still assume that the subject and camera have already been chosen and operate in 2D. The shared limitation across all of these works is that each assumes the other variables are fixed: camera control fixes pose and light; lighting design fixes subject and camera; portrait relighting fixes everything and works in 2D. To our knowledge, prior work has not addressed the joint pre-shutter planning of subject pose, camera configuration, and controllable lighting for prompt-conditioned portrait photography in 3D scenes. We close this gap through the Photographic Scene Graph, which jointly encodes spatial affordances, scene-human relations, and lighting structure in a unified representation, enabling holistic portrait planning.

\section{Methodology}
\begin{figure*}[t]
    \centering
    \includegraphics[width=\textwidth]{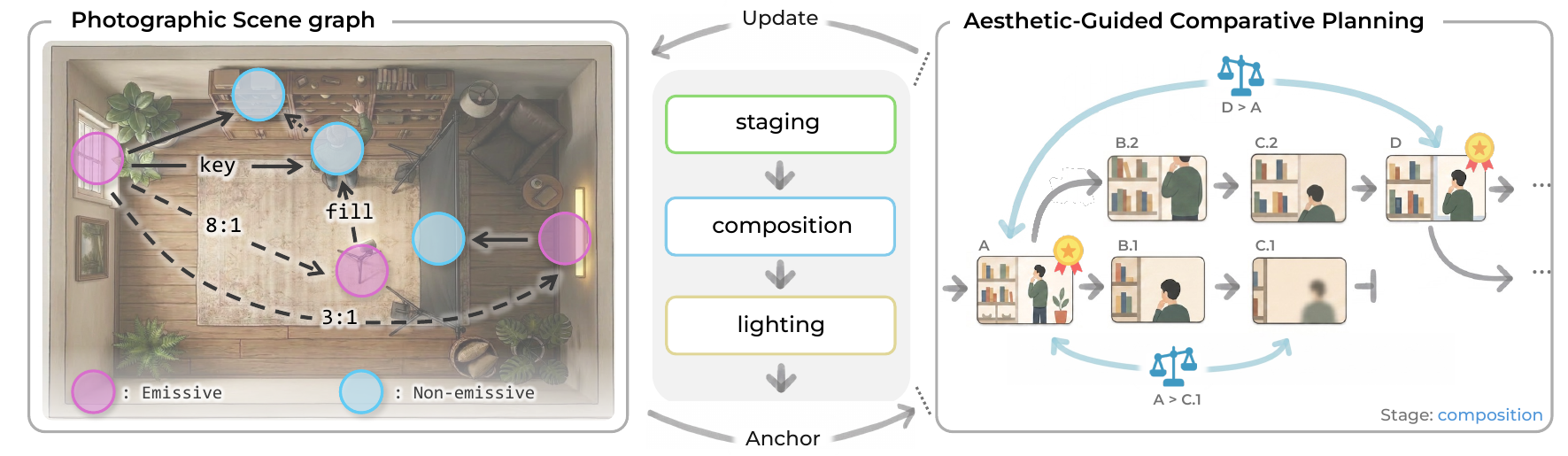}
    \caption{\textbf{Pipeline Overview.} We progressively construct a Photographic Scene Graph to ground
aesthetic-guided comparative planning. Left: the graph represents scene nodes
(e.g., window, lamp, bookshelf), the human subject, controllable lights, and
their spatial and photometric relations. Right: the comparative planning loop,
shown with composition as an example, where the \textsc{Photographer}
iteratively proposes candidate plan states and the \textsc{Judge} compares each new viewfinder observation against the frontier to accept, refine, or reject the state.}
    \label{fig:search}
\end{figure*}

\subsection{Preliminaries}

\paragraph{Scene Representation (\(E\)).}
We consider both the geometry and photometry for the shooting environment and use a unified representation for real-world and virtual scenarios.
First, a global occupancy field
\(\mathcal{O}: \mathbb{R}^3 \rightarrow \{0,1\}\) conceptually encodes rigid scene geometry,
where occupied regions are used primarily to validate shooting actionability. 

For the photometric structure, we require the scene to have a radiance function. A camera ray
with origin \(\mathbf{o}\) and outgoing direction \(\boldsymbol{\omega}_o\) then queries the scene radiance, which we denote as \(L(\mathbf{o}, \boldsymbol{\omega}_o)\). In virtual environments, this radiance is produced by physically-based rendering (PBR).

\paragraph{Human Representation (\(\mathcal{H}\)).}

We represent the subject with SMPL-X. Treating shape, facial expression,
and appearance, garments as fixed inputs, we plan the body \& hand pose
$\boldsymbol{\theta}_b,\boldsymbol{\theta}_h$ and root transform
$(\mathbf{R},\mathbf{t})\in SE(3)$, writing the human state as
$\mathcal{H}=\{\boldsymbol{\theta}_b,\boldsymbol{\theta}_h,
\mathbf{R},\mathbf{t}\}$. This state induces an articulated mesh used
as a geometric proxy for pose planning and body--scene collision checking.

\paragraph{Controllable Lighting Representation (\( \mathcal{L} \)):}
In portrait photography, scene lighting is often supplemented with
controllable lighting equipment, including light sources (e.g., key, fill, and rim lights) and non-emissive occluders such as negative fill. These devices are used to sculpt the subject for
aesthetic purposes~\cite{hunter2015light}. Consequently, we decompose the total
incident radiance \(L_i\) into the environmental contribution \(L_{i,env}\) and
the controllable photographic contribution \(L_{i,ctrl}\):
\begin{equation}
    L_i(\mathbf{x}, \boldsymbol{\omega}_i)
    = L_{i,env}(\mathbf{x}, \boldsymbol{\omega}_i)
    + L_{i,ctrl}(\mathbf{x}, \boldsymbol{\omega}_i; \mathcal{L})
\end{equation}

We parameterize \(\mathcal{L}=\{l_1,\dots,l_{N_l}\}\) as controllable lighting
devices. Each device \(l_{n_l}\) is defined by a rigid transformation in $SE(3)$, and geometric shape and size; emitters additionally include radiometric power and color or temperature.

\paragraph{Camera Representation (\( \mathcal{C} \)) and Image Formation:} 
We adopt a thin-lens camera model that unifies geometric projection with physically-based exposure simulation. The camera is parameterized as \(\mathcal{C} = (\mathbf{T}, \mathbf{K}, \mathbf{P})\), where \(\mathbf{T} \in SE(3)\) defines the camera extrinsic and \(\mathbf{K}\) encodes focal length \(f\) and principal point, together determining the ray origin \(\mathbf{o}\) and direction \(\boldsymbol{\omega}_o\) for each image coordinate.

We use an \emph{aperture-priority} camera, with photographic controls
\(\mathbf{P}=\{N_f,\Delta\}\), where the f-number \(N_f\) controls depth-of-field
and \(\Delta\) denotes exposure compensation in stops relative to the scene-metered baseline. ISO is fixed at 100 for notation simplicity, and
the shutter time $\tau$ is determined by camera metering; we assume an idealized shutter without motion blur. Under this setting, the exposure value  \(\mathrm{EV}_{100}\) is calculated as:
\begin{equation}
    \mathrm{EV}_{100} = \log_2 \frac{N_f^2}{\tau}.
\end{equation}
For each image coordinate \(\mathbf{u} \in \Omega\), the camera model back-projects \(\mathbf{u}\) into a ray \((\mathbf{o}, \boldsymbol{\omega}_o(\mathbf{u}))\); we write \(\mathbf{L}(\mathbf{u}) \equiv L(\mathbf{o}, \boldsymbol{\omega}_o(\mathbf{u}))\) as shorthand for the rendered radiometric signal along that ray. The final image is produced by a \textit{shutter imaging operator} \(\mathcal{S}\), which abstracts the image formed when the planned camera is triggered:
\begin{equation}
    \mathbf{I}(\mathbf{u}) = \mathcal{S}(E, \mathcal{H}, \mathcal{C}, \mathcal{L})\big|_{\mathbf{u}}
    = \Gamma\!\left( \phi\!\left( \kappa\,2^{-(\mathrm{EV}_{100} + \Delta)}\,\mathbf{L}(\mathbf{u}) \right) \right),
\end{equation}
where \(\kappa\) is a calibration constant absorbing lens transmittance, and \(\phi\) is a monotone saturating response before the tone-mapping operator \(\Gamma\), reflecting finite exposure latitude of camera sensors ~\cite{lin2011revisiting,chen2019analyzing}. Overall, this formulation couples geometric composition \((\mathbf{T}, \mathbf{K})\) with photographic control \((N_f, \Delta)\) while leaving \(\mathcal{S}\) as an abstract mapping from scene state to the final image.

\subsection{Problem Formulation}
\label{sec:problem_formulation}
We introduce the task of \textit{aesthetic portrait photography planning}. Given an initially unobserved static scene \(E\), a human subject
\(\mathcal{A}\) providing fixed identity shape and appearance and rigging, and an optional user prompt \(y\) specifying the desired portrait style or constraints, our goal is to generate an aesthetic portrait plan
\[
    s^\ast=(\mathcal{H}^\ast,\mathcal{C}^\ast,\mathcal{L}^\ast),
\]
where \(\mathcal{H}^\ast\) specifies the subject pose and placement,
\(\mathcal{C}^\ast\) specifies the camera configuration, and
\(\mathcal{L}^\ast\) specifies the controllable lighting setup. The plan should satisfy two key constraints:
\begin{enumerate}
    \item \textbf{Actionability}. The plan should be executable in the scene:
    the subject, camera, and lights occupy valid free-space configurations; the human
    body avoids body-scene penetration; and the pose should be stable without floating. The planned exposure should be feasible
    given the scene's photometric structure and the camera's dynamic range.

    \item \textbf{Photographic Aesthetics}. The resulting viewfinder image should be
judged preferable to alternative candidate plans under the prompt. This
preference is assessed in terms of subject-scene coherence, expressive pose and
placement, compositional intent, and lighting design.
\end{enumerate}

\subsection{Aesthetic-Guided Comparative Planning}
To coordinate physical actionability with 2D photographic aesthetics, our key
insight is to use relative aesthetic judgement to guide planning with decoupled roles.
Specifically, our system includes three roles: a \textsc{Photographer}, an
\textsc{Actor}, and a \textsc{Judge}. The \textsc{Photographer} propose edits to plan state, including
camera, lighting \(\mathcal{C},\mathcal{L}\), and staging (i.e., human pose and placement) proposal \(\mathcal{H}\). The \textsc{Actor}
attempts to realize the proposed staging in the environment subject to physical actionability. Both
roles directly manipulate the plan state, resulting in a new viewfinder observation. The \textsc{Judge} does not directly manipulate the plan but instead provides comparative aesthetic judgment on the observations, which drives the planning process.

We denote a planning state as
\(s_t=(\mathcal{H}_t,\mathcal{C}_t,\mathcal{L}_t)\), and its resulting
viewfinder image as \(\mathbf{I}_t=\mathcal{S}(E,s_t)\), where \(t\)
indexes the planning step. During planning, we maintain an aesthetic frontier
\(\mathcal{F}_t=\{(s_i,\mathbf{I}_i,a_i)\}_{i=1}^{K_t}\), where \(s_i=(\mathcal{H}_i,\mathcal{C}_i,\mathcal{L}_i)\) is an accepted best-so-far candidate
state, \(\mathbf{I}_i=\mathcal{S}(E,s_i)\) is its viewfinder observation, and \(a_i\) stores stage and judgement metadata. At each step, the \textsc{Judge} compares the new observation
\(\mathbf{I}_t\) relative to frontier observations
\(\{\mathbf{I}_i\}_{(s_i,\mathbf{I}_i,a_i)\in\mathcal{F}_t}\), and decides based on relative aesthetic judgment whether the new state
\(s_t\) should be added to the frontier, further refined, or discarded. When
refinement is required, the \textsc{Photographer} proposes a revision
\(\Delta s\) to the current plan state and applies it to transition to the next
state \(s_{t+1}\). During staging, this additionally requires the
\textsc{Actor} to realize the revision under physical constraints. This process
iterates until a stopping criterion is met, such as reaching the maximum
planning budget. In this way, we integrate 2D aesthetic preference into a
state-space planning process. Comparative aesthetic guidance directs the search
toward more promising planning directions, as illustrated in
Fig.~\ref{fig:search}. 

\subsection{Photographic Scene Graph}

Planning to improve the aesthetics in $\textbf{I}_t$ is an inverse problem that necessitates understanding the interactions between $\mathcal{H},\mathcal{C},\mathcal{L}$ and the scene \(E\). 
To achieve this, we introduce \textit{Photographic Scene Graph} (Photographic SG). A Photographic SG \(\mathcal{G}\) is a semantic representation of \(E\) and $s_t$, recording both spatial affordances and lighting structure of the scene:
\begin{equation}
    \mathcal{G} =
    (\mathcal{V}_{non}, \mathcal{V}_{emi},
    \mathcal{E}_{n2n}, \mathcal{E}_{e2n}, \mathcal{E}_{e2e}).
\end{equation}
Non-emissive nodes \(\mathcal{V}_{non}\) include both scene objects of interest
(e.g., a chair or a tourist attraction) and body parts of interest (e.g.,
face, torso), storing attributes including affordance tags and reflected-light
metering \(\mathrm{EV}_{100}\). Emissive nodes \(\mathcal{V}_{emi}\) represent principal portrait-relevant light
sources, including dominant environmental emitters and controllable light devices introduced during planning.

We define three types of relations: (1) spatial relations between non-emissive nodes \(\mathcal{E}_{n2n}\), (2)
directed emissive-to-non-emissive light influence \(\mathcal{E}_{e2n}\), and (3) emissive-to-emissive relative source strength \(\mathcal{E}_{e2e}\). For the \(\mathcal{E}_{e2e}\), we let \textsc{Photographer} actively probe the scene $E$ so that it can understand its photometric structure with minimal calibration. Specifically, for each emitter \(v_A\in\mathcal{V}_{emi}\), we estimate its isolated
contribution to a subject by ambient subtraction,
\(M_A^{\Delta}=M_{A+\mathrm{amb}}-M_{\mathrm{amb}}\), where \(M\) denotes the spot-metered luminance of a
Lambertian probe placed at the subject anchor, and \(M_{\mathrm{amb}}\) can be estimated by turning off all controllable lights and occluding dominant environmental emitters. We then record the emitter-to-ambient ratio and pairwise emitter ratio as:
\begin{equation}
    r_{A:\mathrm{amb}}=\frac{M_A^{\Delta}}{M_{\mathrm{amb}}},
    \qquad
    r_{A:B}=\frac{M_A^{\Delta}}{M_B^{\Delta}}
    \left(\frac{d_A}{d_B}\right)^2,
    \quad v_A,v_B\in\mathcal{V}_{emi},
\end{equation}
where \(d_A,d_B\) are light-to-subject distances. This normalization reduces dependence on absolute calibration and provides a practical proxy for the concept of \textit{lighting ratio} in portrait photography~\cite{hunter2015light}.

The \textsc{Photographer} progressively constructs and updates the Photographic
SG \(\mathcal{G}_t\) at each planning step using viewfinder observations, camera
readings, and active probes of viewpoint and lighting. This construction is
semi-automated: the MLLM-based \textsc{Photographer} parses each stepwise
observation into semantic nodes and spatial relations, while reflected-light
metering and photometric relations are algorithmically probed through active
interaction with the 3D scene.

\subsection{SG-Anchored Staging, Composition, and Lighting}
We now describe the stage-wise planning procedure, where the evolving photographic SG
provides shared spatial and photometric context across staging, composition, and
lighting.

\paragraph{Staging.}
Staging decides the subject pose and root transform. Fundamentally, we view this problem as harmonizing \(\mathcal{E}_{n2n}\) between human and scene nodes by considering affordance and aesthetics. \textsc{Photographer} queries
\(\mathcal{G}\) for affordance-compatible non-emissive nodes, such as seats, and uses them as candidate anchors for placing
\(\mathcal{H}\). Each placement proposal is grounded in the image plane, tracked across nearby views, and used to condition a 2D generative prior for proposing staging sketches. To lift these 2D proposals into 3D, we estimate SMPL-X pose \(\boldsymbol{\theta}_b, \boldsymbol{\theta}_h\) from
2D proposals via SMPLer-X~\cite{cai2023smpler}, instantiate the body with the subject-specific shape
\(\boldsymbol{\beta}\) induced by \(\mathcal{A}\), and calculate root transform through multi-view triangulation. The \textsc{Actor} then attempts to realize \(\mathcal{H}\) in \(E\) subject to affordance and physical constraints, and the \textsc{Judge} provides image-level feedback to drive the next staging proposal. 

\paragraph{Composition.} We use the term composition to denote planning of \(\mathcal{C}\). The goal of composition is to organize the spatial relationship among \emph{subjects} within the view frustum. In portraiture, both the human and scene nodes such as landmarks and furniture can serve as subjects. The \textsc{Photographer} therefore queries \(\mathcal{G}\) to perform constrained composition planning. This constraint has two aspects: it controls the visibility of desired nodes \(v\in\mathcal{V}_{non}\) and guides the balance of inter-subject relationships \(\mathcal{E}_{n2n}\) for aesthetic purposes. At each step, the \textsc{Photographer} reviews the viewfinder observation \(\mathbf{I}_t\) against the two constraints and proposes a revision \(\Delta \mathcal{C}\) to the camera parameters.

\paragraph{Lighting.} This stage controls the lighting devices and camera exposure. Conditioned on the user prompt and the \textsc{Judge}’s critique, \textsc{Photographer} first
selects a portrait lighting pattern such as Rembrandt lighting, as
a structured initialization around an anchor node \(v\in\mathcal{V}_{non}\). Specifically, each preset defines a group of light devices with relative position and strength. It then iteratively refines each controllable light's rigid transformation,
size, power, and color temperature, or tries another preset. As in real portrait photography~\cite{hunter2015light}, this
refinement is essential yet tricky: small changes in source position, size, or power can
substantially alter facial modeling, subject separation, shadows, and exposure
under the ambient scene light. The Photographic SG guides this process in
two ways. First, \(\mathcal{E}_{e2n}\) and \(\mathcal{E}_{e2e}\) encode how
existing emitters illuminate the subject anchor and their relative strength. It informs the selection of devices to complement, reshape, or counteract the ambient
contribution. Second, the reflected-light attribute \(\mathrm{EV}_{100}\) on \(v\in\mathcal{V}_{non}\) functions as a semantic meter for the scene, constraining the \textsc{Photographer} to keep the intended subjects within the camera dynamic range and to match the desired tone.

\section{Experiments and Results}

\begin{figure*}[t]
    \centering
    \includegraphics[width=\textwidth]{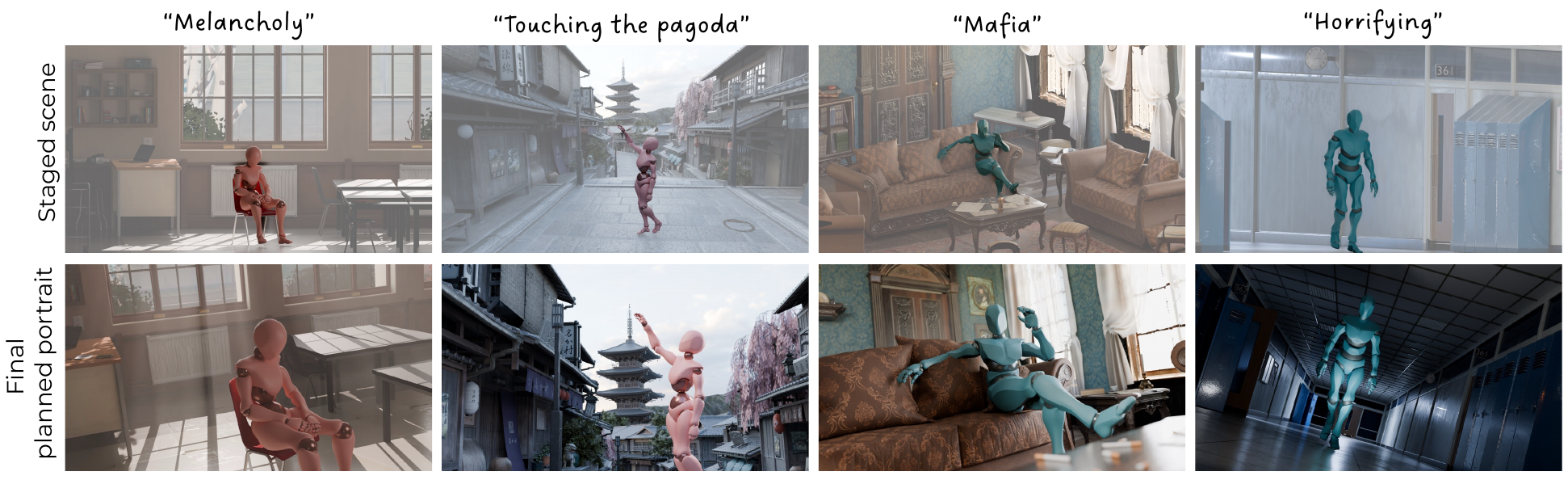}
    \caption{\textbf{Qualitative visualization of our planning approach.} Each column shows the input prompt, the staged 3D scene, and the final shoot under planned camera and lighting control. Best viewed in color and zoomed in.}

    \label{fig:visualization}
\end{figure*}

\subsection{Implementation Details}
For the virtual environment, we use Blender with the Cycles renderer. All assets and lighting setups are configured to be physically plausible, including scale, materials~\cite{burley2012physically}, and light sources. For actor staging, we use Gemini-Flash-2.5-Image (i.e., NanoBanana) for 2D pose proposal, CoTracker~\cite{karaev2024cotracker} for cross-view anchor tracking, and SMPLer-X-H32~\cite{cai2023smpler} for SMPL-X body recovery. For the planning system, we implement \textsc{Photographer} as an MLLM agent by prompting a GPT-5.4-mini model, and the \textsc{Judge} as GPT-5.4 by default.  For a complete planning round, we empirically set the maximum number of steps to 3 for staging and 7 for composition and lighting. We implement the system on a server with a single NVIDIA RTX 4090-D GPU. Per-task runtime ranges from 5--25 minutes and is bounded by rendering complexity. The detailed system parameters are included in the supplementary materials. The full codebase and benchmark will be released upon publication.
\subsection{Evaluation Protocol}

\paragraph{Benchmark Set.}
We construct a benchmark set for aesthetic portrait planning. It contains tasks across \(14\) static 3D scenes, covering 8 indoor and 6 outdoor environments with diverse lighting conditions. We sample a total of \(50\) tasks with diverse user prompts for evaluation. For each case, a method receives the scene, the character, and the prompt, then outputs a complete portrait plan with its associated rendering. 

\paragraph{Baselines.}
To the best of our knowledge, no existing system addresses the joint planning of
human pose, camera configuration, and lighting control in a unified 3D portrait framework.
We therefore compare against baselines that represent alternative degrees of
photographic reasoning and scene grounding. \textit{Random Planner} samples
valid cameras in free space, places a rest-pose human in front of the camera, and uses preset lighting. \textit{Template
Photographer} samples SMPL poses, with eye-level or three-quarter cameras, and
standard portrait lighting as a rule-based lower bound. \textit{Image-Only
Planner} plans iteratively according to viewfinder images and the prompt, without scene graph anchoring.
\textit{Spatial-Graph Planner} uses affordances and spatial relations
without photometric nodes and edges for iterative refinement. \textit{Photographic-Graph One-Pass} uses the full
graph to generate a complete plan in one step, without iterative revision.
\textit{Photographic-Graph Greedy} uses the full graph and refines iteratively, but it exploits local view aesthetic signals without using the aesthetic frontier. We compare these baselines against our full coordinated planner.

\subsection{Metrics}

\paragraph{3D Physical Actionability.}
We evaluate actionability with metrics: skeletal human-scene collision, static
pose balance, and exposure validity. For collision, we report the bone-penetration-free rate
\begin{equation}
    R_{\mathrm{coll}}=\frac{1}{N}\sum_i
    \mathbf{1}\!\left[P_{\mathrm{skel}}(s_i)=0\right],
\end{equation}
where \(\mathcal{B}_s\) denotes skeletal samples of the SMPL-X bones.

For static balance, we report
\begin{equation}
    R_{\mathrm{bal}}=\frac{1}{N}\sum_i S(s_i), \quad
    S(s)=
    \mathbf{1}\!\left[
    \pi_{\perp g}(\mathrm{CoM}_{\mathcal{H}})
    \in \mathrm{ConvHull}(\pi_{\perp g}(\mathcal{Q}_s))
    \right],
\end{equation}
where \(\pi_{\perp g}\) projects points onto the plane perpendicular to gravity,
\(\mathcal{Q}_s=\{\mathbf{x}\in\mathcal{B}^{\mathrm{sup}}_s:
d_{\mathcal{O}}(\mathbf{x})\leq\epsilon_c\}\) are support-contact samples from
load-bearing SMPL-X bones within contact threshold \(\epsilon_c\), and
\(\mathrm{CoM}_{\mathcal{H}}\) is estimated using anthropometric segment-mass
priors~\cite{winter2009biomechanics}.

For exposure, we first compute the valid-pixel fraction
\begin{equation}
    p_{\mathrm{valid}}
    =
    \frac{1}{|\Omega|}
    \sum_{\mathbf{u}\in\Omega}
    \mathbf{1}\!\left[
    2^{-s^-}
    \leq
    \rho(\mathbf{u})
    \leq
    2^{s^+}
    \right],
    \quad
    \rho(\mathbf{u})=\frac{\mathbf{L}(\mathbf{u})}{L_{\mathrm{mid}}},
\end{equation}
where \(L_{\mathrm{mid}}\) is the scene-linear luminance mapped to middle gray
under the planned exposure. We report its smoothed logit:
\begin{equation}
    V_{\mathrm{exp}}
    =
    \log
    \frac{p_{\mathrm{valid}}+10^{-6}}
    {1-p_{\mathrm{valid}}+10^{-6}},
\end{equation}
where \(\Omega\) is the image plane and \(s^-=s^+=3\) characterize empirical reliable signal-exposure latitude of digital cameras~\cite{dxomark2015d7200}.

\paragraph{2D Aesthetic Quality.}

Generic IAA models~\cite{kirstain2023pickapic,xu2023imagereward} are not well matched to
our task: they judge isolated images or prompt-image preference, and do not
explicitly account for whether a portrait uses the given scene, subject, and
use prompt effectively under the judgment subjectivity. We
therefore use a two-alternative forced choice (2AFC)
protocol that assess relatively from 4 dimensions: subject staging, camera composition,
lighting \& exposure, and overall aesthetic quality. Specifically, we perform round-robin tournament comparisons for photo produced by different baselines under same input, then estimate their performance using Bradley--Terry (BT) models~\cite{bradley1952rank}:

\begin{equation}
    \Pr(i \succ j)=\frac{\exp(\beta_i)}{\exp(\beta_i)+\exp(\beta_j)},
\end{equation}
where \(\beta_i\) and \(\beta_j\) denote the latent photographic capabilities of methods
\(i\) and \(j\), respectively.  We report both the point estimate and 95\% confidence interval of \(\beta\).

\renewcommand{\dbltopfraction}{0.95}
\renewcommand{\dblfloatpagefraction}{0.8}
\renewcommand{\textfraction}{0.05}
\setcounter{dbltopnumber}{2}
\newcommand{\btcell}[2]{\makebox[2.7em][r]{\ensuremath{#1}}\ensuremath{{\scriptstyle\pm#2}}}
\newcommand{\btbest}[2]{\makebox[2.7em][r]{\ensuremath{\mathbf{#1}}}\ensuremath{{\scriptstyle\boldsymbol{\pm}\mathbf{#2}}}}

\begin{table*}[!t]
    \centering
    \caption{\textbf{Main quantitative results on the portrait planning benchmark.}
    We report collision-free rate, balance, and exposure validity for actionability.
    Aesthetic columns report Bradley--Terry preference scores with point estimates and
    95\% confidence intervals. All metrics are higher the better.}
    \label{tab:main_quantitative}
    \small
    \begin{tabular}{lcccccccc}
        \toprule
        &
        \multicolumn{3}{c}{Actionability Scores} &
        \multicolumn{5}{c}{Aesthetics (Bradley-Terry $\beta$)} \\
        \cmidrule(lr){2-4}\cmidrule(lr){5-9}
        Method &
        \(R_{\mathrm{coll}}\) \(\uparrow\) &
        \(R_{\mathrm{bal}}\) \(\uparrow\) &
        \(V_{\mathrm{exp}}\) \(\uparrow\) &
        Staging \(\uparrow\) &
        Composition \(\uparrow\) &
        Light \& Exposure \(\uparrow\) &
        Overall (MLLM) \(\uparrow\) &
        Overall (Human) \(\uparrow\) \\
        \midrule
        Random Planner &
        1.00 & 0.00 & 1.08 &
        \btcell{-1.93}{0.27} &
        \btcell{-1.81}{0.24} &
        \btcell{-1.50}{0.25} &
        \btcell{-2.13}{0.25} &
        \btcell{-1.43}{0.20} \\

        Template Photographer &
        0.95 & 0.02 & 0.93 &
        \btcell{-1.15}{0.22} &
        \btcell{-0.25}{0.17} &
        \btcell{-1.06}{0.21} &
        \btcell{-1.71}{0.22} &
        \btcell{-0.72}{0.17} \\

        Image-Only Planner &
        0.84 & 0.18 & 1.32 &
        \btcell{0.47}{0.19} &
        \btcell{-0.10}{0.17} &
        \btcell{0.23}{0.19} &
        \btcell{0.12}{0.17} &
        \btcell{0.15}{0.16} \\

        Spatial-Graph Planner &
        0.94 & 0.58 & 1.49 &
        \btcell{0.80}{0.19} &
        \btcell{0.61}{0.17} &
        \btcell{0.29}{0.20} &
        \btcell{0.84}{0.17} &
        \btcell{0.47}{0.17} \\

        Photographic-Graph One-Pass &
        0.82 & 0.42 & 1.41 &
        \btcell{0.29}{0.19} &
        \btcell{0.19}{0.17} &
        \btcell{0.19}{0.19} &
        \btcell{0.50}{0.17} &
        \btcell{0.01}{0.17} \\

        Photographic-Graph Greedy &
        0.80 & 0.52 & 1.35 &
        \btcell{0.59}{0.19} &
        \btcell{0.64}{0.17} &
        \btcell{0.61}{0.20} &
        \btcell{1.07}{0.18} &
        \btcell{0.54}{0.17} \\

        \textbf{Ours Full} &
        0.92 & 0.56 & 1.56 &
        \btbest{0.92}{0.19} &
        \btbest{0.72}{0.17} &
        \btbest{1.25}{0.23} &
        \btbest{1.30}{0.18} &
        \btbest{0.96}{0.19} \\
        \bottomrule
    \end{tabular}
\end{table*}

We collect judgement from both human and an MLLM (Gemini-3-pro). Expert
raters (n=19) contribute 13140 valid annotations, with inter-annotator agreement (Cohen's kappa) of 0.66, 0.61, 0.69, 0.73 on each dimension. We also report
zero-shot MLLM judgments as a secondary automatic evaluator, motivated by recent
findings that MLLMs produces better human-aligned zero-shot aesthetic
judgments~\cite{jiang2025mllmaesthetics} under 2AFC setting, compared with IAA models. In our experiment, the MLLM alignment (Spearman's rho) with human judgment is 0.60, 0.75, 0.92, and 0.78 on each dimension, justifying its use as a complementary evaluator.

\begin{figure*}[!htbp]
  \centering
  \includegraphics[width=\linewidth,keepaspectratio]{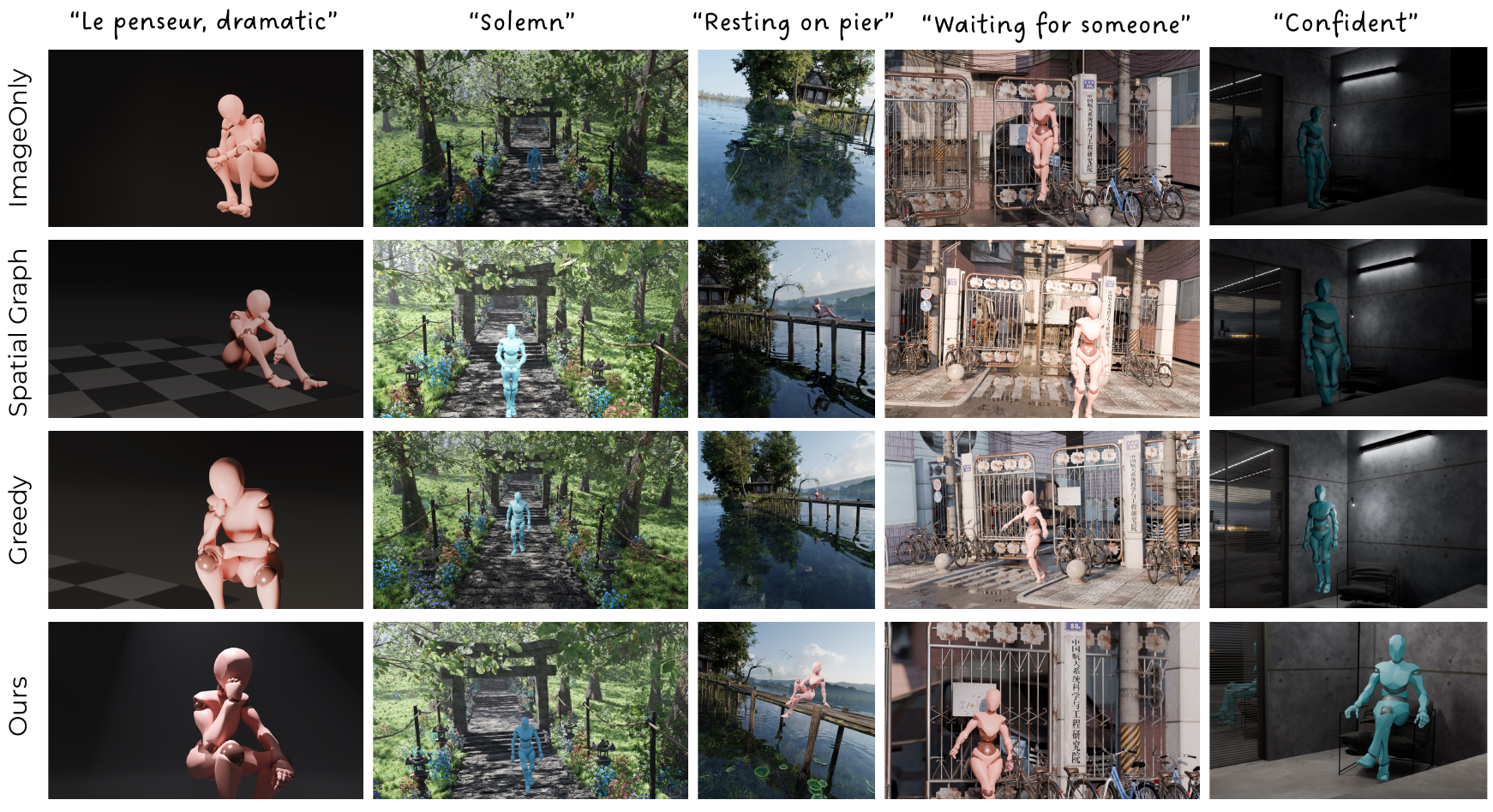}
  \caption{\textbf{Qualitative comparison of our method and baselines.} Compared with ours, baselines less faithfully coordinate pose, camera, and lighting to match the prompt. The generated pose can be floating or awkward, the composition can be unbalanced or even miss the subject(s), and the lighting can be indiscriminately flat, harsh, or underexposed, failing to match the desired tone implied by prompt. Zoom in for details.}
  \label{fig:qual_comp}
\end{figure*}

\begin{figure*}[!htbp]
  \centering
  \includegraphics[width=\linewidth,keepaspectratio]{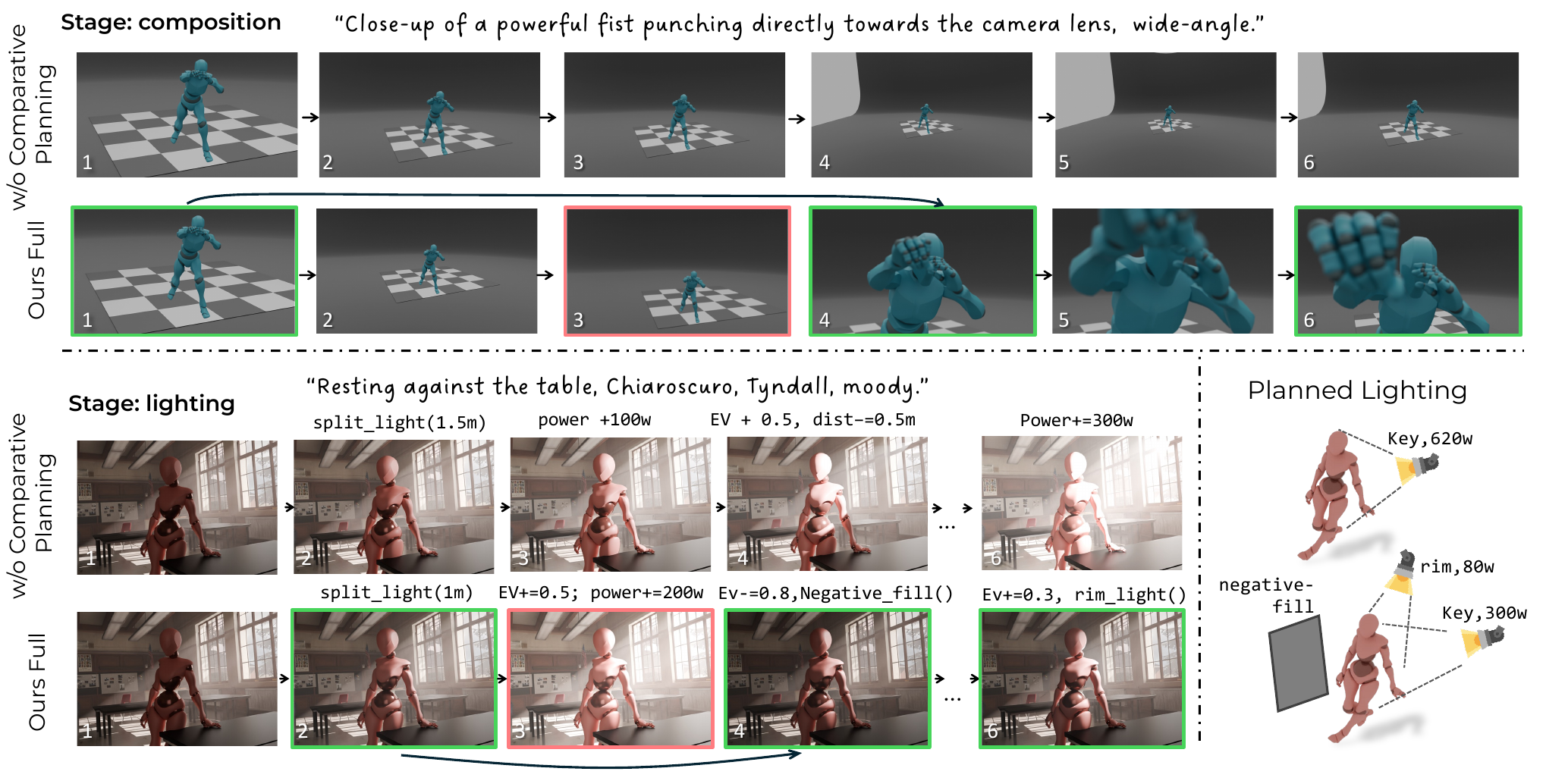}
  \caption{\textbf{Ablation of Comparative Planning.} Comparative planning allows the planner to revert to earlier frontier states when a refinement direction is judged to degrade aesthetics. In the top example, it helps to solve the ambiguity between wide-angle and camera distance. In the bottom example, it informs the planner to use a negative fill on the composite side to enhance the contrast instead of continuously strengthening the key light. Green outline: accepted frontier state.}
  \label{fig:stagewise_ablation}
\end{figure*}

\subsection{Quantitative Results}

\paragraph{Main Results.} Tab.~\ref{tab:main_quantitative} reports the final-plan performance across
actionability and aesthetic preference. Our full method obtains the highest
overall preference under both evaluators, with BT scores of
\(1.30\pm0.18\) from the MLLM judge and \(0.96\pm0.19\) from human raters, while maintaining high physical actionability. SG-based methods show a substantial improvement over the Image-only baselines: the scene graph constrains planning to be scene-aware and helps produce affordance-grounded human poses (higher \(R_{\mathrm{coll}}\), \(R_{\mathrm{bal}}\)) and manageable exposure settings (higher \(V_{\mathrm{exp}}\)). On the other hand, iterative planning is also essential: the Photographic One-pass method, which uses the full Photographic SG but without iterative refinement, performs worse than the iterative-planning-based Spatial-Graph Planner. We attribute this to the lack of interaction with the scene for grounding the plan. Comparative planning built on top of the iterative planner substantially improves both actionability and preference scores, as demonstrated by the comparison between our full method and Photographic-Graph Greedy.

\paragraph{Stage-wise Ablations.}
To isolate the contributions of comparative planning (\(\mathcal{F}\) for short) and the Photographic SG \(\mathcal{G}\), we checkpoint the plan at the end of each stage of our full method and fork it into different branches with either module removed. Tab.~\ref{tab:stagewise_ablation} summarizes the ablation-internal scores using the MLLM evaluator. Adding \(\mathcal{G}\) allows the planner to make informed adjustments to camera and lighting. However, this does not readily improve planning quality in the absence of \(\mathcal{F}\). This is because \(\mathcal{G}\) and its induced constraints are progressively constructed during planning. We empirically find that planning without \(\mathcal{F}\) is prone to self-reinforcing state dynamics. This observation is reminiscent of cycling and local lock-in in memoryless local
search in traditional control theory literature~\cite{glover1989tabu,mayne2000constrained}. We visualize this phenomenon in Fig.~\ref{fig:stagewise_ablation}. Combining both \(\mathcal{F}\) and \(\mathcal{G}\) achieves the best result by performing scene- and state-aware planning.

\begin{table}[!htbp]
    \centering
    \caption{\textbf{Stage-wise ablation of Comparative Planning ($\mathcal{F}$) and Photographic SG ($\mathcal{G}$)}. no-op means skip the stage. Higher is better.}
    \label{tab:stagewise_ablation}
    \small
    \begin{tabular}{lccc}
        \toprule
        Method & Staging $\beta$ \(\uparrow\) & Comp. $\beta$ \(\uparrow\) & Light $\beta$ \(\uparrow\) \\
        \midrule
        no-op & -- & \btcell{-0.39}{0.26} &         \btcell{-0.31}{0.25}  \\
        \(+\mathcal{F}\) & \btbest{-0.25}{0.27} & \btcell{-0.02}{0.24} &         \btcell{-0.27}{0.25}  \\
        \(+\mathcal{G}\) & \btbest{-0.05}{0.26} & \btcell{-0.14}{0.24} &         \btcell{0.12}{0.25}  \\
        \(+(\mathcal{F}, \mathcal{G})\) & \btbest{0.29}{0.27} & \btbest{0.55}{0.24} & \btbest{0.46}{0.24} \\
        \bottomrule
    \end{tabular}
\end{table}

\subsection{Qualitative Results and Visualization}

\begin{figure*}[p]
  \centering
  \includegraphics[width=\linewidth,keepaspectratio]{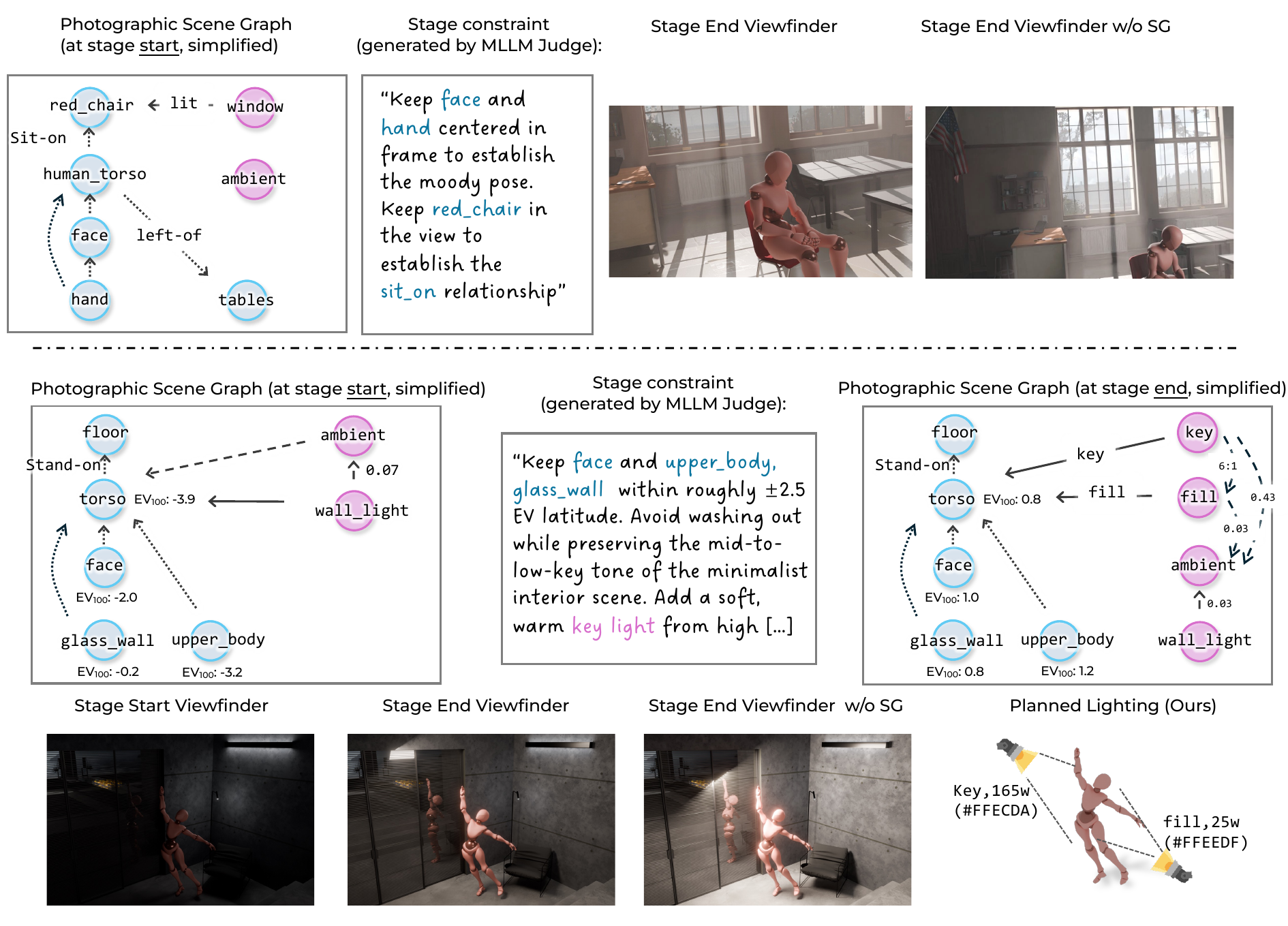}
  \caption{\textbf{Visualization of Photographic Scene Graph.} Top: An example of SG-anchored composition, prompt:  ``\textit{Melancholy}''. The MLLM judge generated constraint to guide the composition to preserve the human pose and the red chair to establish the mood. Bottom: An example of SG-anchored lighting, prompt: ``\textit{Gracefully dancing near the glass}''. The SG provide photometric structure of the scene, which is used to guide the lighting design and exposure planning. Simplified means we manually select the relevant nodes and edges for visualization readability. Zoom in for details.}
  \label{fig:sg}
\end{figure*}

Fig.~\ref{fig:visualization} visualizes the prompt, staged scene, and
the final captured portrait for four sample tasks. The proposed method produces aesthetic plans that coordinate the human subject and scene, composed with cinematic camera angles and lighting. Fig.~\ref{fig:qual_comp} compares the result produced by our method with selected baselines. The results show that our method produces more visually appealing portraits that better harmonize the scene and subject and adhere better to the prompt. Fig.~\ref{fig:sg} visualizes of how the Photographic SG anchors planning in composition and lighting stage.

\subsection{Limitations and Future Work}
In this work, the staging is limited to the body and hand pose, while leaving the important facial expression and gaze direction fixed. This omission is due to two reasons: (1) Technically, intricate facial expressions make it challenging to transfer expression between 2D guidance and an arbitrary 3D articulated human asset at high fidelity. (2) Evaluation of facial expression is more challenging. In practice we tested human assets with expressions and find users suffer from uncanny valley effects when rendered realistically, which can bias the judgment~\cite{tinwell2011facial}. This is also why we only utilized the mannequin asset in our experiments. We therefore leave facial-expression planning as a future direction, which may require joint learning with an expression prior and use a more specialized evaluation protocol. 

\section{Conclusion}
We introduced aesthetic portrait planning in 3D scenes, a task that shifts computational portraiture from post-capture 2D editing to pre-capture 3D decision making. Our method combines a Photographic SG with aesthetic-guided comparative planning to jointly search over pose, camera, and lighting while respecting physical feasibility in the scene. This formulation allows us to solve the inverse problem of mapping 2D portrait aesthetics to 3D actionable plans through scene-grounded interaction and observation. Experiments show that the proposed method improves both actionability and human preference over competitive baselines. We hope this setting opens new opportunities for scene-aware photographic assistance, virtual production, and embodied creative planning.

\clearpage

\clearpage

\bibliographystyle{ACM-Reference-Format}
\bibliography{main}

\end{document}